\RequirePackage{fix-cm}
\documentclass[smallextended]{svjour3}       
\smartqed  
\def\makeheadbox{\relax}
\usepackage{graphicx}
\usepackage{mathptmx}      
\usepackage[utf8]{inputenc}
\usepackage{physics}
\usepackage{dsfont}
\usepackage[amssymb]{SIunits}
\newcommand{\rmi}{\ensuremath{\mathrm{i}}}
\newcommand{\rme}{\ensuremath{\mathrm{e}}}
\newcommand{\rmd}{\ensuremath{\mathrm{d}}}
\newcommand{\bleq}{\ensuremath{\mathrel{\phantom{=}}}}
\newcommand{\nnl}{\nonumber\\}
\newcommand{\schr}{Schr{\"o}\-din\-ger}
\renewcommand{\vec}[1]{\mathbf{#1}}
\newcommand{\reals}{\mathds{R}}
\newcommand{\complex}{\mathds{C}}
\usepackage{color}

\begin{document}

\title{Is gravitational entanglement evidence for the quantization of spacetime?}


\author{M. Kemal Döner         \and
        André Großardt
}


\institute{Friedrich Schiller University Jena\\
Institute for Theoretical Physics,
Fröbelstieg 1,
07743 Jena, Germany\\
\email{kemal.doner@uni-jena.de \and andre.grossardt@uni-jena.de}
}

\date{}

\maketitle

\begin{abstract}
Experiments witnessing the entanglement between two particles interacting only via the gravitational field have been proposed as a test whether gravity must be quantized. In the language of quantum information, a non-quantum gravitational force would be modeled by local operations with classical communication (LOCC), which cannot generate entanglement in an initially unentangled state. This idea is criticized as too constraining on possible alternatives to quantum gravity. We present a parametrized model for the gravitational interaction of quantum matter on a classical spacetime, inspired by the de Broglie--Bohm formulation of quantum mechanics, which results in entanglement and thereby provides an explicit counterexample to the claim that only a quantized gravitational field possesses this capability.
\end{abstract}

\section{Introduction}\label{sec:intro}
Contemporary physics, at its most fundamental level, is in a somewhat peculiar situation, relying on two separate and apparently incompatible theories. On one hand, there is matter, described by the Fock space states of interacting quantum fields or, more specifically, the Standard Model with its 12 elementary fermions and its $SU(3) \times SU(2) \times U(1)$ gauge symmetry, from which nonrelativistic, quantum mechanical behavior follows, at least in principle. On the other hand, there is spacetime, a Lorentzian 4-manifold which provides the metric and differential structures with which dynamical laws for matter can be defined and whose curvature is determined by the matter distribution via Einstein's equations.

The quest of ``quantum gravity'', in its broadest meaning, refers to the goal of finding \emph{some} common mathematical framework which, in the appropriate limits of observed physical phenomena, can embed the predictions of both quantum and gravitational physics.
By virtue of the largely different mathematical structures, the prevailing believe is that this must be achieved by \emph{some} sort of ``quantization'' of gravity, for instance in the sense of promoting some objects in the theory of general relativity (the metric, curvature, connection, volume, or area elements, \dots) to a Hilbert space structure, albeit the precise meaning remains obscure and differs from model to model.

Contrariwise, one may pose the question what would need to change about the formalism of \emph{quantum physics} in order to be compatible with the principles of general relativity---a ``gravitization of quantum mechanics'' in the words of Penrose~\cite{penroseGravitizationQuantumMechanics2014}---such that quantum matter could be consistently described on a (classical) spacetime manifold, including its backreaction on spacetime. Leaving aside (important) mathematical details, the dynamics of quantum fields on a curved spacetime can be formulated as a well-defined theory~\cite{waldQuantumFieldTheory1994,barQuantumFieldTheory2009}. The more fundamental challenge is the opposite question: how does one model the effect of quantum matter on spacetime curvature?

The most straightforward approach to model this coupling of quantum matter to classical gravity is via the semiclassical Einstein equations~\cite{mollerTheoriesRelativistesGravitation1962,rosenfeldQuantizationFields1963}
\begin{equation}\label{eqn:sce}
R_{\mu\nu} - \frac{1}{2} g_{\mu\nu} R = \frac{8 \pi G}{c^4} \bra{\Psi} \hat{T}_{\mu\nu} \ket{\Psi} \,,
\end{equation}
where the left-hand side is the Einstein tensor, constructed from the scalar and tensor curvatures $R$ and $R_{\mu\nu}$ as well as the metric $g_{\mu\nu}$, and the right-hand side contains the expectation value of the stress-energy operator in the quantum state $\Psi$. Its consistency as a fundamental model has been the subject of discussions~\cite{eppleyNecessityQuantizingGravitational1977,kibbleSemiClassicalTheoryGravity1981,pageIndirectEvidenceQuantum1981,mattinglyWhyEppleyHannah2006,albersMeasurementAnalysisQuantum2008}, although with no conclusive result. In nonrelativistic situations, it can be understood~\cite{bahramiSchrodingerNewtonEquationIts2014} as resulting in a \emph{wave function dependent} Newtonian gravitational potential
\begin{equation}\label{eqn:vsc}
V_\text{sc}(t,\vec r_1, \vec r_2) = -\frac{G m_1 m_2}{\abs{\vec r_1 - \ev{\vec r_2}}} 
-\frac{G m_1 m_2}{\abs{\ev{\vec r_1} - \vec r_2}}
+ V^{(1)}_\text{self}(t,\vec r_1)
+ V^{(2)}_\text{self}(t,\vec r_2) 
\end{equation}
for two particles of masses $m_1$ and $m_2$, where $\ev{\cdot}$ denotes the expectation value in the two-particle state with spatial wave function $\Psi(t,\vec r_1,\vec r_2)$ and
\begin{equation}
V^{(i)}_\text{self}(t,\vec r_i) = - G m_i^2 \int \rmd^3 r_1' \,\rmd^3 r_2' \,
\frac{\abs{\Psi(t,\vec r_1',\vec r_2')}^2}{\abs{\vec r_i - \vec r_i'}}
\end{equation}
is the self-gravitational potential of the $i$-th particle. This self-gravitational attraction of even a single particle has previously been considered as a route towards experimentally testing semiclassical gravity~\cite{carlipQuantumGravityNecessary2008,giuliniGravitationallyInducedInhibitions2011,yangMacroscopicQuantumMechanics2013,grossardtOptomechanicalTestSchrodingerNewton2016}.
Recently, experiments have been proposed~\cite{boseSpinEntanglementWitness2017b,marlettoGravitationallyInducedEntanglement2017} which would be looking for the difference between the first two mutual interaction terms in the semiclassical potential~\eqref{eqn:vsc} as compared to the potential expected from perturbatively quantized gravity:
\begin{equation}\label{eqn:vqg}
 V_\text{qg}(\vec r_1, \vec r_2) = -\frac{G m_1 m_2}{\abs{\vec r_1 - \vec r_2}} \,.
\end{equation}
Quantized gravity then predicts an entangled two-particle state, whereas the semiclassical model would leave an initially separable state unentangled.

The claim made by the proponents of these tests, however, goes beyond distinguishing the two potentials~\eqref{eqn:vsc} and~\eqref{eqn:vqg}, stating instead that it is ``impossible'' for two particles to develop entanglement from a classical field~\cite{marshmanLocalityEntanglementTabletop2020b} and that ``anything capable of entangling two quantum systems and satisfying locality (plus a few other assumptions) must itself be quantum.''~\cite{marlettoAnswersFewQuestions2019}
Therefore, so the idea, experimental evidence for entanglement would not only rule out the potential~\eqref{eqn:vsc} but \emph{any} semiclassical model for gravity.
This idea has been criticized by Hall and Reginatto~\cite{hallTwoRecentProposals2018} who point out that entanglement is strictly prohibited only for the specific type of classical interactions as introduced by Koopman~\cite{koopmanHamiltonianSystemsTransformation1931} and show that in the hybrid model of quantum-classical ensembles~\cite{hallInteractingClassicalQuantum2005} entanglement can increase. Similarly,
Pal et al.~\cite{palExperimentalLocalisationQuantum2021} show that entanglement can occur between two initially unentangled qubits through a third qubit whose reduced density matrix remains diagonal for the whole experiment, \emph{if} there is some initial entanglement already present in the entire three-qubit system. 

Here, we present a different model as a counterexample to the claim that gravitational entanglement is evidence against semiclassical theories; one that makes use of the trajectory\footnote{This ``trajectory approach'' is briefly mentioned, but not elaborated upon, also by Hall and Reginatto~\cite{hallTwoRecentProposals2018}. Another brief consideration~\cite{andersenQuantumStatisticsBohmian2019} of the spin entanglement proposals~\cite{boseSpinEntanglementWitness2017b,marlettoGravitationallyInducedEntanglement2017} from the perspective of Bohmian trajectories was brought to our attention shortly before finalizing this work.} as an additional ``hidden'' variable in the de Broglie--Bohm theory, and is closer in spirit to the mean-field approach based on the semiclassical Einstein equations~\eqref{eqn:sce}. Contrary to the mean-field approach, a dependence of the gravitational field on these hidden variables allows it to possess information about the outcome of measurements beyond the classical information encoded in expectation values. Furthermore, it allows for the definition of hybrid models in which the gravitational potential depends on both the trajectories and the wave function, and which thereby interpolate between the maximally entangling model with a point particle source and the non-entangling semiclassical potential~\eqref{eqn:vsc}.

The structure of this paper is as follows:
In the next section~\ref{sec:bohmian}, we present explicitly how the Bohmian trajectories can be used to source a Newtonian gravitational potential with the correct classical limit, and predicting entanglement between two gravitationally interacting particles. We review how, in the de Broglie--Bohm picture, entanglement of localized particles can be understood as arising from additional local fields on physical 3-space as the effect of conditional potentials~\cite{norsenTheoryExclusivelyLocal2010,norsenCanWaveFunction2015}, and we discuss the implications for semiclassical gravity. Section~\ref{sec:interpol} provides an example how this approach can be generalized to a class of models, with the mean-field potential~\eqref{eqn:vsc} and the trajectory based model from section~\ref{sec:bohmian} as limiting cases. In section~\ref{sec:witness} we explicitly calculate the spin-entanglement witnesses~\cite{boseSpinEntanglementWitness2017b,guffOptimalFidelityWitnesses2021} proposed for non-classicality tests of gravity, and show how they would confirm entanglement for the semiclassical models presented. The discussion section~\ref{sec:discussion} reviews the implications of our results for the interpretation of experimental tests of gravitationally induced entanglement and addresses important limitations.

\section{Gravity sourced along Bohmian trajectories}\label{sec:bohmian}

Consider a number of quantum particles with nonrelativistic energies, whose gravitational interaction we would like to describe within the frameworks of both the de Broglie--Bohm approach to quantum theory and classical general relativity. Gravity, according to general relativity, is modeled by the curvature of a classical spacetime manifold---which in nonrelativistic situations is fully determined by the Newtonian potential to good approximation.

In addition to classical spacetime and the wave function, we introduce particle coordinates $\vec q_i$, which in the de Broglie--Bohm theory satisfy a guiding equation,
\begin{equation}\label{eqn:guiding}
\frac{\rmd \vec q_i(t)}{\rmd t}
= \frac{\hbar}{m_i} 
\left.\Im\left(\frac{\nabla_i \Psi(t;\vec r_1,\dots,\vec r_N)}{\Psi(t;\vec r_1,\dots,\vec r_N)}\right)\right\vert_{\vec r_1 = \vec q_1(t), \dots, \vec r_N = \vec q_N(t)} \,,
\end{equation}
depending on the $N$-particle wave function $\Psi$. Having these trajectories $\vec q_i(t)$ at our disposal, enables us to use them as point-particle sources for a Newtonian gravitational interaction.
In the Schrödinger equation for $N=2$ particles we then introduce the potential 
\begin{equation}\label{eqn:vbb}
V_\text{bb}(\vec r_1, \vec r_2, \vec q_1, \vec q_2) = 
-\frac{G m_1 m_2}{\abs{\vec r_1 - \vec q_2}}
-\frac{G m_1 m_2}{\abs{\vec q_1 - \vec r_2}}
+\gamma_0(\vec q_1,\vec q_2)\,,
\end{equation}
such that the two dynamical equations form a coupled system. Let us study the consequences of this potential, starting with its classical limit. We notice that the \schr\ equation yields the usual Ehrenfest theorem
\begin{equation}
m_i \frac{\rmd^2}{\rmd t^2}\ev{\vec r_i} 
 = -\ev{\nabla_i V_\text{bb}(\vec r_1, \vec r_2, \vec q_1, \vec q_2)} \,,
\end{equation}
which results in the same classical equations of motion for the expectation values as the quantum potential~\eqref{eqn:vqg}. Whereas for the quantum potential a single term determines the motion of both particles, for the potential~\eqref{eqn:vbb} the first term determines the motion of the first particle in the gravitational field of the second, the second term the motion of the second particle in the field of the first. The final term $\gamma_0$ in equation~\eqref{eqn:vbb} is a function of only the Bohmian trajectories $\vec q_i$ and does not contribute to the classical limit. It is necessary in order to maintain consistency with the experimentally confirmed gravitational phase shift~\cite{colellaObservationGravitationallyInduced1975,fixlerAtomInterferometerMeasurement2007,nesvizhevskyMeasurementQuantumStates2003}.

For the further analysis, we turn to the local formulation of the de Broglie--Bohm theory~\cite{norsenTheoryExclusivelyLocal2010,norsenCanWaveFunction2015}, defined via the conditional wave functions
\begin{equation}
	\begin{split}
		\psi_1(t,x) &\equiv \Psi(t;x,q_2(t)) \\
		\psi_2(t,x) &\equiv \Psi(t;q_1(t),x) \,.
	\end{split}
\end{equation}
We consider only one spatial dimension for the subsequent discussion, with the generalization to three dimensions being straightforward. The guiding equations~\eqref{eqn:guiding} then can be written in terms of these conditional wave functions,
\begin{equation}\label{eqn:cwf-guiding}
 \frac{\rmd q_i(t)}{\rmd t} = \frac{\hbar}{m_i}
 \left.\Im\left(\frac{\partial \psi_i(t,x)/\partial x}{\psi_i(t,x)}\right)\right\vert_{x = q_i(t)} \,,
\end{equation}
implying that the evolution for the $i$-th particle depends only on the conditional wave function $\psi_i$. The evolution of $\psi_i$ itself, however, depends nontrivially on both the other particle coordinates and the remaining conditional wave functions, thereby allowing for the quantum mechanical entanglement.
This is expressed through the $n$-th order entanglement potential fields
\begin{equation}\label{eqn:entanglement-fields}
\begin{split}
\Pi_1^{(n)}(t,x) &\equiv \frac{1}{\psi_1(t,x)}\, \left.\frac{\partial^n \Psi(x,x_2,t)}{\partial x_2^n}\right|_{x_2=q_2(t)} \\
\Pi_2^{(n)}(t,x) &\equiv \frac{1}{\psi_2(t,x)}\, \left.\frac{\partial^n \Psi(x_1,x,t)}{\partial x_1^n}\right|_{x_1=q_1(t)} \,.
\end{split}
\end{equation}
The conditional wave functions then satisfy the \schr\ equations
\begin{equation}
 \rmi \hbar \frac{\partial}{\partial t} \psi_i(t,x)
 = -\frac{\hbar^2}{2 m_i} \frac{\partial^2}{\partial x^2} \psi_i(t,x) + V_i^\text{eff}(t,x) \psi_i(t,x)
 \label{eqn:schr-cwf}
\end{equation}
 with the time dependent effective potentials
\begin{subequations}\label{eqn:eff-pot}\begin{align}
 V_1^\text{eff} &= \gamma_0(q_1,q_2)
 -\frac{G m_1 m_2}{\abs{q_1 - q_2}}
 -\frac{G m_1 m_2}{\abs{x - q_2}}
 +\rmi \hbar \frac{\rmd q_2}{\rmd t}\,\Pi_1^{(1)}
 -\frac{\hbar^2}{2 m_2}\,\Pi_1^{(2)} \\
V_2^\text{eff} &= \gamma_0(q_1,q_2)
 -\frac{G m_1 m_2}{\abs{q_1 - q_2}}
 -\frac{G m_1 m_2}{\abs{q_1 - x}}
 +\rmi \hbar \frac{\rmd q_1}{\rmd t}\,\Pi_2^{(1)}
 -\frac{\hbar^2}{2 m_1}\,\Pi_2^{(2)} \,.
\end{align}\end{subequations}
The potential fields $\Pi_i^{(n)}$ depend, of course, on the conditional wave functions. For a solution consistent with the full 2-particle \schr\ equation, each potential field must itself obey a partial differential equation depending on higher order entanglement fields~\cite{norsenCanWaveFunction2015}. The locality of the \schr\ equation~\eqref{eqn:schr-cwf} comes at the price of this dependency on an infinite number of entanglement fields, which can be truncated at some order $n$ for an approximate treatment, capturing entanglement up to a certain degree. For our purpose, this approach has its main advantage in how it explicitly reveals the entanglement.

For any \emph{given} solution $\Psi$ of the full 2-particle \schr\ equation, one can calculate the four potential fields of first and second order and treat them as given external potentials. Equations~\eqref{eqn:schr-cwf} together with the guiding equations~\eqref{eqn:cwf-guiding} then describes a system of equations which is coupled only through the dependence of the effective potentials~\eqref{eqn:eff-pot} on the trajectories $q_i(t)$. Of course, knowing the full solution $\Psi$ the solutions for the $\psi_i$ and $q_i$ follow immediately. In this sense, this system is of little use for the purpose of \emph{solving} the dynamical equations. It can, however, be useful in order to \emph{analyze} the dynamical properties of the system.

The trajectories described by equation~\eqref{eqn:cwf-guiding} generally diverge from each other in a way determined by the spreading of the wave function. In a \emph{classical} situation, where the wave function remains sharply peaked over the relevant time period, the trajectories, therefore, remain close to the classical trajectories $u_i(t) = \ev{x_i}$, which solve the classical equations of motion
\begin{equation}
 m_i\,\frac{\rmd^2}{\rmd t^2} u_i(t)
 = \frac{\rmd}{\rmd x_i}\left.\frac{G m_1 m_2}{\abs{x_1 - x_2}}\right\vert_{x_1 = u_1(t),\,x_2 = u_2(t)} \,.
\end{equation}
Approximating $q_i(t) \approx u_i(t)$ in equations~\eqref{eqn:eff-pot} rather than using the guiding equation, one obtains from~\eqref{eqn:schr-cwf} two fully decoupled \schr\ equations for the conditional wave functions $\psi_i$. As ususal, to lowest semiclassical order the effective potentials result in a phase
\begin{equation}
 \phi_i^\text{eff} = -\frac{1}{\hbar} \int_0^\tau \rmd t \, V_i^\text{eff}(t,u_i(t))
\end{equation}
after time $\tau$.
Consider an intially separable wave function,
\begin{equation}
\Psi(0;x_1,x_2) = \alpha(x_1)\beta(x_2)\,,
\end{equation}
such that
\begin{alignat}{2}
\psi_1(0,x) &= \alpha(x)\beta(q_2)
&\qquad
\psi_2(0,x) &= \alpha(q_1)\beta(x) \nnl
\Pi_1^{(1)}(0,x) &= \frac{\beta'(q_2)}{\beta(q_2)}
&\qquad
\Pi_2^{(1)}(0,x) &= \frac{\alpha'(q_1)}{\alpha(q_1)} \\
\Pi_1^{(2)}(0,x) &= \frac{\beta''(q_2)}{\beta(q_2)}
&\qquad
\Pi_2^{(2)}(0,x) &= \frac{\alpha''(q_1)}{\alpha(q_1)}
\nonumber \,,
\end{alignat}
and, choosing $\gamma_0(q_1,q_2) = G m_1 m_2 \abs{q_1 - q_2}^{-1}$ as well as assuming Gaussian wave functions of widths $\sigma_i$ peaked at $x_1 = u_1$ and $x_2 = u_2$, respectively,
\begin{equation}
 V_i^\text{eff}(0,u_i) = 
 -\frac{G m_1 m_2}{\abs{u_1 - u_2}}
 + V_i^\text{ent} \,,
\end{equation}
with the entanglement potentials
\begin{equation}
 V_1^\text{ent} = \frac{\hbar^2}{2 m_2 \sigma_2^2}
 \qquad\text{and}\qquad
 V_2^\text{ent} = \frac{\hbar^2}{2 m_1 \sigma_1^2} \,.
\end{equation}
These entanglement potentials only result in a trajectory-independent phase, whereas the gravitational potential yields a phase
\begin{equation}
 \phi_\text{grav} = \frac{G m_1 m_2\, \tau}{\hbar \,\Delta u} 
\end{equation}
for each conditional wave function, where we assume a constant separation $\Delta u = \abs{u_1(t) - u_2(t)}$. This identical phase of $\psi_1$ and $\psi_2$ can be interpreted as a phase of the 2-particle wave function $\Psi$. Note that it is identical to the phase predicted from the quantum potential~\eqref{eqn:vqg} and used in reference~\cite{boseSpinEntanglementWitness2017b}, \emph{if} the function $\gamma_0$ is chosen as the positive gravitational energy between the two Bohmian particle locations, i.\,e. if the first two terms in equations~\eqref{eqn:eff-pot} cancel. For other choices it differs, specifically by a factor of two when choosing $\gamma_0 \equiv 0$.

Thus far, we only considered classical states with a well defined trajectory. Adressing the experimental situation~\cite{boseSpinEntanglementWitness2017b} of a superposition of two classical trajectories for each of the two particles, $u_{1,2}^\pm$, we have the initially separable wave function
\begin{equation}
\Psi(0;x_1,x_2) = \frac{1}{2} \left(\alpha(x_1 - u_1^+) + \alpha(x_1 - u_1^-)\right) \left(\beta(x_2 - u_2^+) + \beta(x_2 - u_2^-)\right) \,,
\end{equation}
with $\alpha$ and $\beta$ symmetric functions sharply peaked around zero. Then the conditional wave functions, which generally depend on the other particle's trajectory, are the trajectory independent
\begin{equation}
\begin{split}
\psi_1(0,x) &= \frac{\beta(0) + \beta(\Delta u)}{2} \left(\alpha(x-u_1^+) + \alpha(x-u_1^-)\right)  \\
\psi_2(0,x) &= \frac{\alpha(0) + \alpha(\Delta u)}{2} \left(\beta(x-u_2^+) + \beta(x-u_2^-)\right) \,,
\end{split}
\end{equation}
both describing a superposition of two classical trajectories. For a given potential $V_1^\text{eff}$, each of the two trajectories in $\psi_1$ acquires its own phase, and accordingly for $\psi_2$. However, the effective potentials $V_i^\text{eff}$ explicitly depend on the \emph{other} particle's trajectory. Therefore, each of the four possible combinations $(u_1^\pm,u_2^\pm)$ acquires a different phase depending on the specific combination of trajectories, resulting in the same phases $\phi_\text{grav}^{++}$, $\phi_\text{grav}^{+-}$, $\phi_\text{grav}^{-+}$, and $\phi_\text{grav}^{--}$ as predicted from the quantum potential~\eqref{eqn:vqg}.

We conclude that the potentials~\eqref{eqn:vbb} and~\eqref{eqn:vqg} make the same predictions for both the classical limit and the gravitational phase shift; in other words, they agree with respect to experimentally tested gravitational phenomena, including the yet untested gravitational entanglement~\cite{boseSpinEntanglementWitness2017b,marlettoGravitationallyInducedEntanglement2017}. Nonetheless, the de Broglie--Bohm inspired model has a semiclassical interpretation in which curvature of a classical spacetime is sourced by the trajectories $\vec q_i(t)$ of the particles. One may argue that this is pure semantics and that a model that makes the same physical predictions is, for all practical purposes, equivalent to quantized gravity. This is why, in the next section~\ref{sec:interpol}, we opt for a hybrid potential that interpolates between the potentials \eqref{eqn:vsc} and \eqref{eqn:vqg} by integrating the modulus-squared of the wave function only about a radius $R$ around the particle coordinates $\vec q_i$.

The local description provides us with an intuitive understanding of the origins of entanglement. In the limit of weak entanglement, starting with initially separable states, the entanglement resulting from the potential fields $\Pi_i^{(n)}$ is negligible compared to that resulting from the interaction potential. A typical quantum potential $V(x_1,x_2)$ results in an effective potential $V_1^\text{eff}(x) = V(x,q_2)$ for the first particle, depending on the second particle's trajectory, and vice versa. The particles become entangled due to this dependency on the other trajectory. It is then evident, that the same entanglement can be achieved if instead of the 2-particle interaction $V(x_1,x_2)$, the 2-particle wave function $\Psi$ experiences a potential that already has an explicit dependence on the trajectories, such as our potential~\eqref{eqn:vbb}.

\section{Mean-field trajectory hybrid model}\label{sec:interpol}
In the previous section we presented a model, based on the de Broglie--Bohm trajectories, in which gravity can be understood as curvature of a single classical spacetime, despite inducing entanglement between two particles. This model presents a counterexample against taking such gravitational entanglement as evidence for a quantized gravitational field. Nonetheless, one may be tempted to argue that this model should be considered a quantum theory in some sense of the word, that it does not have a consistent relativistic generalization, or disqualify it as a legitimate counterexample for some other reason.
In order to make the relation to mean-field semiclassical gravity more visible, we introduce a hybrid potential that interpolates between the potentials \eqref{eqn:vsc} and \eqref{eqn:vbb} by integrating the modulus-squared of the wave function only about a radius $R$ around the particle coordinates $\vec q_i$. It, therefore, describes an entire class of models, parametrized by $R$, which include mean-field semiclassical gravity as a limiting case. We show how the entanglement between the particles decreases as $R$ grows larger, reaching no entanglement only in the limit $R \to \infty$ of the semiclassical Einstein equations.

For introducing our semiclassical model, we begin with the most general $N$-particle Schrödinger equation
\begin{align}\label{eqn:schroedinger}
\rmi \hbar \frac{\partial \Psi(t;\vec r_1,\dots,\vec r_N)}{\partial t}
&= \Bigg( -\frac{\hbar^2}{2} \sum_{i=1}^N \frac{\nabla_i^2}{m_i} 
\nnl &\bleq + V(t;\vec r_1,\dots,\vec r_N;\vec q_1,\dots,\vec q_N;\Psi)\Bigg) \Psi(t;\vec r_1,\dots,\vec r_N) \,,
\end{align}
where $\nabla_i$ is the gradient with respect to the coordinate $\vec r_i$, and the potential $V$, besides an explicit dependence on time and position coordinates, has both a functional dependence on the wave function and depends on the Bohmian particle coordinates $\vec q_i(t)$ which are determined by the guiding equation~\eqref{eqn:guiding}.

Equations~\eqref{eqn:schroedinger} and \eqref{eqn:guiding} form a coupled nonlinear system, which can in principle be solved for both the wave function solutions $\Psi$ and the particle trajectories $\vec q_i$.
We specify the potential to take the following form, depending on the parameter $R$:
\begin{subequations}\begin{align}
V_R(\vec r_1,\dots,\vec r_N;\vec q_1,\dots,\vec q_N;\Psi)
&= -G \sum_{i=1}^N \sum_{j=1}^N m_i m_j \frac{1 - \delta_{ij} f_\text{reg}(R)}{N_j(t,R)} \nnl
&\bleq \times \int \rmd^3 r \frac{\chi(R,\vec q_j - \vec r) P_j(t,\vec r)}{\abs{\vec r_i - \vec r}} \nnl
&\bleq + \gamma_R(\vec q_1,\dots,\vec q_N)\label{eqn:vr}
\intertext{with}
\chi(R,\vec r) &= \begin{cases}
1 & \text{for } \abs{\vec r} \leq R \\
0 & \text{for } \abs{\vec r} > R \end{cases} \\
P_j(t,\vec r) &= \int \rmd^3 r_1 \cdots \int \rmd^3 r_{j-1} \int \rmd^3 r_{j+1} \cdots \int \rmd^3 r_N \nnl &\bleq \times
\abs{\Psi(t;\vec r_1,\dots,\vec r_{j-1},\vec r,\vec r_{j+1},\dots,\vec r_N)}^2 \\
N_j(t,R) &= \int \rmd^3 r \, P_j(t,\vec r) \chi(R,\vec q_j - \vec r) \,,
\intertext{a regularization function $f_\text{reg} : \reals_+ \to [0,1]$ with}
f_\text{reg}(R) &\to \begin{cases} 0 & \text{for } R \to \infty \\ 1 & \text{for } R \to 0 \quad\text{``sufficiently fast'',}\end{cases}
\end{align}\end{subequations}
and $\gamma_R$ a, for now, arbitrary function of only the $\vec q_i$ with $\gamma_R \to 0$ for $R \to \infty$.
$P_i$ is simply the marginal probability distribution for the $i$-th particle. $\chi$ is the characteristic function for a sphere of radius $R$, limiting the integration to a spherical region around the particle positions; the functions $N_i$ ensure normalization.
$V_R$ has no explicit time dependence but is implicitly time dependent through the coordinates and wave function. Additional linear potentials can be straightforwardly added to the \schr\ equation~\eqref{eqn:schroedinger}. The regularization function $f_\text{reg}$ is required, such that no divergent self-interaction terms appear in the limit $R \to 0$; its precise form is irrelevant for the further discussion, as self-gravitational effects will be neglected.

Evidently, the potential $V_R$ mimics the behavior of the Bohmian potential~\eqref{eqn:vbb}---and thus the quantum potential~\eqref{eqn:vqg}---in the limit $R \to 0$. In this limit, the wave function dependence of $V_R$ vanishes. In the limit $R \to \infty$, on the other hand, $V_R$ turns into the semiclassical potential~\eqref{eqn:vsc}, in which case it no longer depends on the Bohmian trajectories $\vec q_i$. Having both the semiclassical, nonlinear coupling to the wave function and the Bohmian trajectories at our disposal, we can source the gravitational potential by a mass distribution associated with $\abs{\Psi}^2$, as in the semiclassical model, with the distinction that only that part of the wave function contributes which lies within a radius $R$ of the actual particle position, as determined by $\vec q_i$.

\subsection{Classical limit}
The \schr\ equation~\eqref{eqn:schroedinger} yields the usual equations of motion
\begin{align}
 \partial_t^2 \ev{\vec r_i} &= -\frac{1}{m_i}
 \ev{\nabla_i V_R(t;\vec r_1, \dots,\vec r_N; \vec q_1,\dots,\vec q_N;\Psi)} \nnl
 &= G \sum_{j=1}^N m_j \frac{1 - \delta_{ij} f_\text{reg}(R)}{N_j(t,R)} \int \rmd^3 r \, \chi(R,\vec q_j - \vec r) P_j(t,\vec r) \ev{\frac{\vec r - \vec r_i}{\abs{\vec r_i - \vec r}^3}}
\end{align}
for the position expectation values. Note that the function $\gamma_R$ does not appear. In the limit $R \to 0$, the potential limits to equation~\eqref{eqn:vbb} and the classical equations of motion follow as derived in section~\ref{sec:bohmian}. For finite $R$ we introduce the characteristic function $\overline{\chi}$ of the complement of the sphere of radius $R$ such that $\chi + \overline{\chi} \equiv 1$. We then have
\begin{subequations}\begin{align}
 \partial_t^2 \ev{\vec r_i} &= 
 \sum_{\substack{j=1\\j\not=i}}^N \left( \vec a_{ij} + \Delta \vec a_{ij} \right)
 + \vec a_i^\text{self}
 \intertext{with}
 \vec a_{ij} &= \frac{G m_j}{N_j(t,R)} \iint \rmd^3 r \, \rmd^3 r' \, P_j(t,\vec r) P_i(t,\vec r') \frac{\vec r - \vec r'}{\abs{\vec r - \vec r'}^3}  \\
 \Delta \vec a_{ij} &= -\frac{G m_j}{N_j(t,R)} \iint \rmd^3 r \, \rmd^3 r' \, \overline{\chi}(R,\vec q_j - \vec r) P_j(t,\vec r) P_i(t,\vec r') \frac{\vec r - \vec r'}{\abs{\vec r - \vec r'}^3}  \\
 \vec a_i^\text{self} &= - G m_i \frac{1 - f_\text{reg}(R)}{N_i(t,R)} \iint \rmd^3 r \, \rmd^3 r' \, \overline{\chi}(R,\vec q_i - \vec r) P_i(t,\vec r) P_i(t,\vec r') \frac{\vec r - \vec r'}{\abs{\vec r - \vec r'}^3} \,. \label{eqn:self-force}
\end{align}\end{subequations}
Note that inside the sphere of radius $R$, where $\chi \equiv 1$, the self-force integral~\eqref{eqn:self-force} vanishes due to the antisymmetry under $\vec r \leftrightarrow \vec r'$.

For quasi-classical particles, localized around $\vec q_i \approx \ev{\vec r_i}$ with a spatial extent far smaller than $R$, we have $P_j(t,\vec r) \approx \delta(\vec r - \vec q_j)$. The integrals via $\overline{\chi}$ outside the radius $R$ then yield negligible contributions, $\Delta \vec a_{ij} \approx 0 \approx \vec a_i^\text{self}$. Furthermore, $N_j(t,R) \approx 1$, and we find the classical Newtonian equations of motion
\begin{align}
 \partial_t^2 \vec q_i &\approx 
 G \sum_{\substack{j=1\\j\not=i}}^N m_j \frac{\vec q_j - \vec q_i}{\abs{\vec q_j - \vec q_i}^3} \,.
\end{align}

\subsection{Two equal mass particles in a double Stern--Gerlach experiment}\label{sec:sg}

\begin{figure}
	\centering
	\includegraphics[scale=0.55]{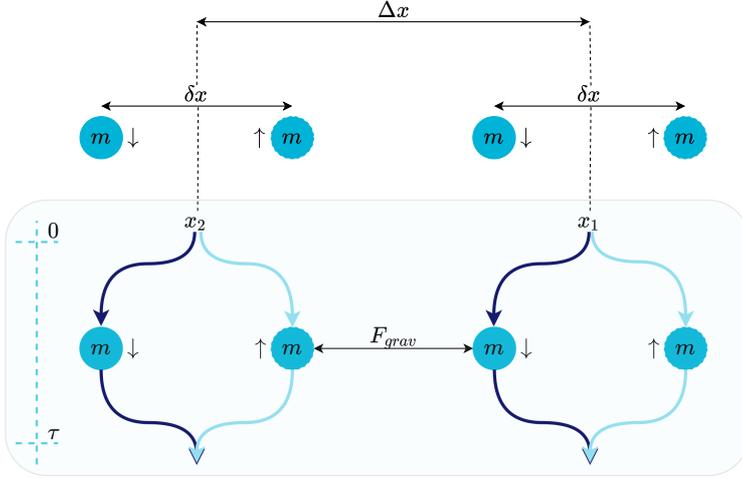}
	\caption{Schematic depiction of the double Stern--Gerlach experiment as proposed by Bose et al.~\cite{boseSpinEntanglementWitness2017b}. Two spin-$\tfrac12$ particles are each brought into superposition of two possible trajectories in an inhomogeneous magnetic field, resulting in a total of four possible trajectory combinations located around $x$ coordinates $u_{1,2}^{s_{1,2}} = \pm (\Delta x/2) + s_{1,2} (\delta x/2)$ depending on the spin eigenvalues $s_{1,2} = \pm 1$.}
	\label{fig:superpositions}
\end{figure}

For the further discussion, we consider the double Stern--Gerlach experiment proposed by Bose et al.~\cite{boseSpinEntanglementWitness2017b} and depicted in figure~\ref{fig:superpositions}. We focus on the special case of two particles of equal mass $m$. Ignoring the self-gravitational terms, that affect only each particle at its site but not both together, we find
\begin{align}\label{eqn:vr-2-part-full}
V_R(\vec r_1,\vec r_2;\vec q_1,\vec q_2;\Psi)
&\approx - \frac{G m^2}{N_1(t,R)} \int \rmd^3 r \frac{\chi(R,\vec q_1 - \vec r) P_1(t,\vec r)}{\abs{\vec r_2 - \vec r}}
\nnl &\bleq - \frac{G m^2}{N_2(t,R)} \int \rmd^3 r \frac{\chi(R,\vec q_2 - \vec r) P_2(t,\vec r)}{\abs{\vec r_1 - \vec r}} \nnl
&\bleq + \gamma_R(\vec q_1,\vec q_2) \,.
\end{align}
We consider a situation where the quasi-classical trajectories for two spin-$\tfrac12$ particles are split in a magnetic field gradient over a short time period $\tau_a$ and recombined after some free flight time $\tau$. Assuming $\tau \gg \tau_a$, we can neglect the gravitational effects during the acceleration period, and only consider the four classical trajectories $\vec u_{1,2}^{s_{1,2}} = (u_{1,2}^{s_{1,2}},0,v t)$ for $t \in [0,\tau]$, split along the $x$-axis with $u_{1,2}^{s_{1,2}} = \pm (\Delta x/2) + s_{1,2} (\delta x/2)$ depending on the spin eigenvalues $s_{1,2} = \pm 1$. 

\begin{figure}
	\centering
	\includegraphics[scale=0.55]{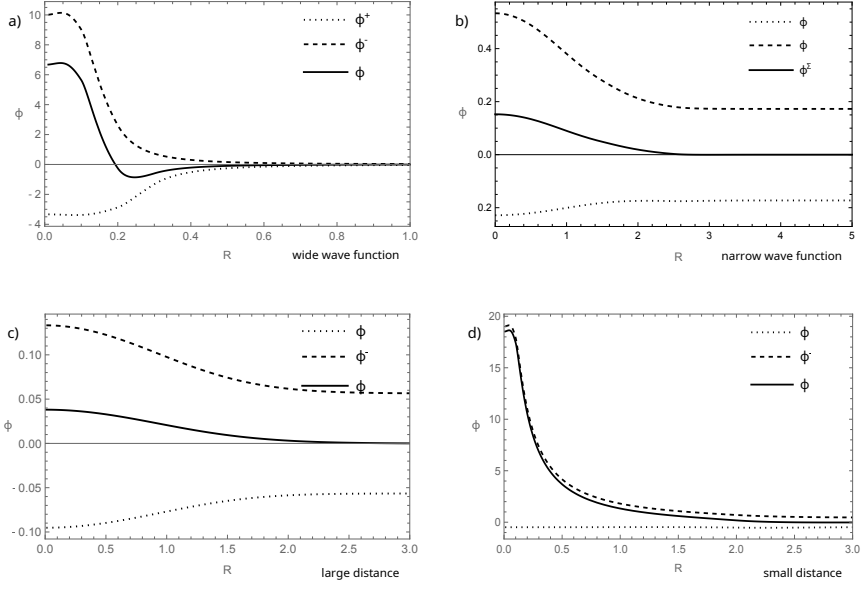}
	\caption{Change of phases $\phi_R^+$, $\phi_R^-$, $\phi_R^{\Sigma}$ as functions of R for different situations; a) wide wave function: $\Delta x=0.25$, $\delta x=0.1$, b) narrow wave function: $\Delta x=2.5$, $\delta x=1$, c) large distance: $\Delta x=3$, $\delta x=0.5$, d) small distance:$\Delta x=2$, $\delta x=1.9$, according to equations~\eqref{eqn:phi-plusminus} and~\eqref{eqn:phase_ave}.
	\label{fig:phase plots}}
\end{figure}

Due to the gravitational attraction, and to lowest order semiclassical approximation, the particles acquire spin-dependent phases obtained by integrating the potential $V_R$ along the classical trajectories (cf. the Appendix for a detailed discussion):
\begin{equation}\label{eqn:spin-dep-phase}
 \phi_R^{s_1s_2} \approx -\frac{1}{\hbar} \int_0^\tau \rmd t \,
 V_R(\vec u_1^{s_1}(t),\vec u_2^{s_2}(t);\vec q_1^{s_1}(t),\vec q_2^{s_2}(t);\Psi(t;\vec u_1^{s_1}(t),\vec u_2^{s_2}(t))) \,.
\end{equation}
These phases still depend nonlinearly on the solution $\Psi$ of the \schr\ equation. As long as the gravitational effects are weak, however, the wave function in equation~\eqref{eqn:spin-dep-phase} can be approximated by the solution of the free \schr\ equation~\cite{grossardtDephasingInhibitionSpin2021a}. For the proper choice of parameters---$m \, \sigma^2 \gg \hbar\,\tau$, where $\sigma$ is the initial width of the wave function---we can also ignore the free spreading and, therefore, consider only the time independent wave function $\Psi(\vec r_1,\vec r_2)$ representing the superposition of all four possible spin combinations. In the case where the particles follow quasi-classical trajectories, the Bohmian trajectories follow closely, $\vec q_i(t) \approx \vec u_i^{s_i}(t)$. The time dependence can then be transformed away or omitted entirely by choosing $v = 0$, resulting in
\begin{equation}
 \phi_R^{s_1s_2} \approx 
 \Gamma \int \rmd^3 r \left( \frac{\chi(R,\vec u_1^{s_1} - \vec r) P_1(\vec r)}{N_1(R) \, \abs{\vec u_2^{s_2} - \vec r}} 
 + \frac{\chi(R,\vec u_2^{s_2} - \vec r) P_2(\vec r)}{N_2(R) \, \abs{\vec u_1^{s_1} - \vec r}}\right)
 + \phi_\gamma^{s_1s_2} \,,
\end{equation}
with the constant $\Gamma = G m^2 \tau/\hbar$ of the dimension of a length.
We ignore the phase contribution $\phi_\gamma^{s_1s_2}$ of $\gamma_R$ for now, choose length units in which the width of wave packets around the trajectories is of order unity, and consider a Gaussian wave function
\begin{align}
\Psi(x_1,r_1,x_2,r_2) &= \sum_{s_1} \sum_{s_2}
\frac{\exp\left[-\frac{1}{2}\left(r_1^2+r_2^2 + (x_1-u_1^{s_1})^2 + (x_2-u_2^{s_2})^2\right)\right]}{2 \sqrt{\pi^3} \left(1 + \exp\left(-\frac{\delta x^2}{4}\right)\right)}
\end{align}
in cylindrical coordinates, $(x,y,z) = (x,r\,\cos\theta,r\,\sin\theta)$. Each of the four contributions to $\Psi$ is spherically symmetric with respect to the corresponding trajectory. We have
\begin{subequations}\begin{align}
  P_i(x,r) &= \frac{\rme^{-r^2}}{2 \sqrt{\pi^3}}
  Q(x-u_i^+,x-u_i^-)
 \intertext{with}
 Q(p,q) &= \frac{\rme^{-p^2} + \rme^{-q^2}}{1 + \exp\left(-\frac{\delta x^2}{4}\right)}
 +\frac{2 \exp\left(-\left(\frac{p + q}{2}\right)^2\right)}{1 + \exp\left(\frac{\delta x^2}{4}\right)}
 \intertext{and the normalization}
 N(R) &= \left(1 - \rme^{-R^2}\right)
 \Bigg[ \frac{\mathrm{erf}\left(R+\frac{\delta x}{2}\right)+\mathrm{erf}\left(R-\frac{\delta x}{2}\right)}{2 \left(1 + \exp\left(\frac{\delta x^2}{4}\right)\right)}
 \nnl &\bleq
 + \frac{\mathrm{erf}(R+\delta x)
 + \mathrm{erf}(R-\delta x)
 + 2 \mathrm{erf}(R)}{4 \left(1 + \exp\left(-\frac{\delta x^2}{4}\right)\right)}
 \Bigg] \,,
\end{align}\end{subequations}
where $N_1(R) = N_2(R) = N(R)$ regardless of the trajectory.
Defining
\begin{equation}
J_R(\xi) = \frac{2}{\sqrt{\pi}} \int_0^R \rmd r 
\frac{r\, \rme^{-r^2}}{\sqrt{r^2 + \xi^2}} 
= \rme^{\xi^2} \left[
\mathrm{erf}\left(\sqrt{R^2 + \xi^2}\right)
- \mathrm{erf}\left(\sqrt{\xi^2}\right)
\right]
\end{equation}
and writing $\Delta u^{s_1s_2} = u_1^{s_1}-u_2^{s_2}$, using $Q(p,q) = Q(q,p) = Q(-p,-q)$, we find
\begin{align}
\phi_R^{s_1s_2} &\approx 
\frac{\Gamma}{N(R)} \int_{-R}^R \rmd x \int_0^R \rmd r \left(
\frac{2 \pi r\,P_1(x + u_1^{s_1},r)}{\sqrt{(x + \Delta u^{s_1s_2})^2 + r^2}}
+ \frac{2 \pi r\,P_2(x + u_2^{s_2},r)}{\sqrt{(x - \Delta u^{s_1s_2})^2 + r^2}} 
\right)
\nnl
&= \frac{\Gamma}{2N(R)} \int_{-R}^R \rmd x
\Bigg(
 Q(x + u_1^{s_1} - u_1^+,x + u_1^{s_1} - u_1^-)
\nnl &\bleq
+ Q(x - u_2^{s_2} + u_2^+,x - u_2^{s_2} + u_2^-)
\Bigg) J_R(x + \Delta u^{s_1s_2}) \,.
\end{align}
Considering the four spin combinations independently, we find a global phase $\Phi_R = \phi_R^{++} = \phi_R^{--}$, as well as the relative phases
\begin{align}\label{eqn:phi-plusminus}
\phi_R^\pm = \phi_R^{\pm\mp} - \Phi_R &\approx 
\frac{\Gamma}{N(R)} \int_{-R}^R \rmd x \,  Q\left(x,x+\delta x\right)
\nnl &\bleq \times \left(
J_R(x \pm \Delta x + \delta x)
- \frac{J_R(x + \Delta x) + J_R(x - \Delta x)}{2}
\right)
\end{align}
and their average
\begin{align}\label{eqn:phase_ave}
\phi_R^\Sigma &= \frac{\phi_R^+ + \phi_R^-}{2} = \frac{\phi_R^{+-} + \phi_R^{-+}}{2} - \Phi_R 
\nnl &\approx 
\frac{\Gamma}{2N(R)} \int_{-R}^R \rmd x \,  Q\left(x,x+\delta x\right)
\Big(
J_R(x + \Delta x + \delta x)
+ J_R(x - \Delta x + \delta x)
\nnl &\bleq 
- J_R(x + \Delta x) - J_R(x - \Delta x)
\Big) \,.
\end{align}
The difference $2 \phi_R^\Delta = \phi_R^+ - \phi_R^-$ is always nonzero and can, therefore, be tuned to take any value by adjusting the prefactor $\Gamma$.
Figure~\ref{fig:phase plots} shows the phases as a function of $R$ for different relative distances $\Delta x$ and $\delta x$ with respect to the wave function width $\sigma$.

The additional phase contribution from a nonzero $\gamma_R$ is
\begin{equation}
 \phi_\gamma^{s_1s_2} = -\frac{\tau}{\hbar} \gamma_R(\vec u_1^{s_1},\vec u_2^{s_2}) \,.
\end{equation}
Assuming that $\gamma_R(\vec q_1,\vec q_2) = \gamma_R(\abs{\vec q_1 - \vec q_2})$ is a function of relative distance only, we have $\phi_\gamma^{++} = \phi_\gamma^{--}$ contributing only to the global phase $\Phi_R$. The relevant phase contributions then are
\begin{equation}
 \phi_\gamma^\pm = \frac{\tau}{\hbar} \left( \gamma_R(\Delta x)
 - \gamma_R(\Delta x \pm \delta x)\right)\,.
\end{equation}

\subsection{Small $R$ expansion}

In the limit $R \to 0$, both the integral from $-R$ to $R$ and the normalization function $N(R)$ tend to zero like $R^3$. Three-fold application of l'Hôpital's rule yields the phases
\begin{equation}\label{eqn:phase-r0}
 \phi_0^\pm = \frac{2 \Gamma}{\abs{\Delta x \pm \delta x}} - \frac{2 \Gamma}{\abs{\Delta x}} + \phi_\gamma^\pm \,.
\end{equation}
We can generalize this to an expansion around small $R \ll 1$, by approximating up to and including $\order{R^5}$:
\begin{align}
 J_R(\xi) &\approx \frac{R^2 }{\sqrt{\pi \xi^2}} \left[1 - \frac{R^2}{2} \left(1 + \frac{1}{2 \xi^2}\right)\right] \\
 N(R) &\approx \frac{R^3\,\rme^{-\delta x^2}}{\sqrt{\pi} \left(1 + \rme^{-\frac{\delta x^2}{4}}\right)} \left[
 \left(1 + \rme^{\frac{\delta x^2}{2}}\right)^2
 \left(1 - \frac{5 R^2}{6}\right)
 + \frac{R^2 \, \delta x^2}{3}
 \left(2 + \rme^{\frac{\delta x^2}{2}}\right)
 \right] \,.
\end{align}
Therefore,
\begin{equation}
 \frac{J_R(\xi)}{N(R)} \approx 
 \frac{\left(1 + \rme^{-\frac{\delta x^2}{4}}\right)}{\left(1 + \rme^{-\frac{\delta x^2}{2}}\right)^2} \left[ \frac{1}{R \abs{\xi}} 
 + \frac{R}{3 \abs{\xi}} \left(
 1 - \frac{3}{4 \xi^2} - \delta x^2 \frac{2 + \rme^{\frac{\delta x^2}{2}}}{\left(1 + \rme^{\frac{\delta x^2}{2}}\right)^2}
 \right) \right] \,.
\end{equation}
For arbitrary functions $q(x)$, $f(x)$, $g(x)$, we have to cubic order in $R$:
\begin{multline}
 \int_{-R}^R \, q(x) \left(\frac{f(x + \xi)}{R} + R g(x + \xi)\right) \rmd x
\approx 2 q(0) f(\xi) \\ + \frac{R^2}{3} \left(
 q(0) \left(6 g(\xi) + f''(\xi)\right)
 + 2 q'(0) f'(\xi) + q''(0) f(\xi)
\right) \,,
\end{multline}
and hence, with $q(x) = Q(x,x+\delta x)$,
\begin{equation}
I_R(\xi) = \frac{\Gamma}{2} \, \int_{-R}^R \rmd x \,  q(x)
\frac{J_R(x + \xi)}{N(R)}
\approx \frac{\Gamma}{\abs{\xi}} \left[1 + \frac{R^2}{12 \xi^2} \left(1 
+ \frac{8 \xi \,\delta x}{1 + \rme^{\frac{\delta x^2}{2}}} \right)\right] \,.
\end{equation}
The phases are then
\begin{align}\label{eqn:phase_r}
\phi_R^\pm &= 2 I_R(\delta x \pm \Delta x) - I_R(\Delta x) - I_R(-\Delta x) + \phi_\gamma^\pm \,.
\end{align}
To lowest order, we again obtain the phases~\eqref{eqn:phase-r0}.
As expected, this is the phase obtained from the Bohmian potential~\eqref{eqn:vbb}, as discussed in section~\ref{sec:bohmian}. Without the function $\gamma_R$, it is twice the phase expected from 
quantum gravity~\cite{boseSpinEntanglementWitness2017b}, although it can be easily amended to recover the quantum result in the limit $R \to 0$ by choosing $\gamma_R(\xi) \to G m^2 I_0(\xi)$ in this limit. Since the prediction of entanglement is based solely on these phases, we expect to be able to witness the same entanglement as for quantum gravity. The maximum amount of entanglement is independent of the choice of $\gamma_R$, only requiring an appropriate rescaling of the parameter $\Gamma$ via the particle mass $m$ and the flight time $\tau$.

\subsection{Large $R$ expansion}

In the limit $R \to \infty$, using the asymptotic expansion of the error function,
\begin{equation}
 \mathrm{erf}(x) = 1 - \frac{\rme^{-x^2}}{\sqrt{\pi}\,x}
 \left(1 - \frac{1}{2 x^2} + \frac{3}{4 x^4} 
 - \frac{15}{8 x^6}
 +\dots \right)\,,
\end{equation}
we find $N(R) \to 1$ and
\begin{equation}
 J_\infty(\xi) = \rme^{\xi^2} \left(1 - \mathrm{erf}(\abs{\xi})\right)\,.
\end{equation}
In order to expand the solution for large $R$ we notice that approximately
\begin{align}
 \frac{J_R(\xi)}{N(R)} \approx J_\infty(\xi) \left(1 + \frac{\rme^{-R^2} R}{2\sqrt{\pi}}
 \delta x^2 \frac{1+2 \rme^{\delta x^2/4}}{1+\rme^{\delta x^2/4}}
 \right)
 - \frac{\rme^{-R^2}}{\sqrt{\pi} R}
 \left(
 1 -\frac{\xi^2}{2 R^2}\right)\,,
\end{align}
where only the $\xi$-dependent part of $J_R(\xi)$ contributes to the phases~\eqref{eqn:phi-plusminus}. With
\begin{equation}
h^\pm(x) = Q(x,x+\delta x) \left(
J_\infty(x \pm \Delta x + \delta x)
- \frac{J_\infty(x + \Delta x) + J_\infty(x - \Delta x)}{2}
\right)
\end{equation}
we find
\begin{align}
\int_{-R}^R h^\pm(x) &=
\int_{-\infty}^\infty h^\pm(x)
- \int_R^\infty \left(h^\pm(x) + h^\pm(-x)\right)
\nnl &\approx \frac{\phi_\infty^\pm}{\Gamma}
+ \frac{\delta x \pm \Delta x}{1 + \rme^{-\delta x^2/4}} 
\frac{\rme^{-R^2}}{\sqrt{\pi} R^3} \,,
\end{align}
and hence, assuming $\phi_\gamma^\pm \to 0$ sufficiently fast,
\begin{align}
\phi_R^\pm &\approx 
 \Gamma \left(1 + \frac{\rme^{-R^2} R}{2\sqrt{\pi}}
 \delta x^2 \frac{1+2 \rme^{\delta x^2/4}}{1+\rme^{\delta x^2/4}}
 \right) \int_{-R}^R \rmd x \, h^\pm(x)
 \pm \frac{\Gamma \, \delta x \, \Delta x}{\rme^{R^2} R^3} \nnl
 &\approx 
\left(1 + \frac{\rme^{-R^2} R}{2\sqrt{\pi}}
 \delta x^2 \frac{1+2 \rme^{\delta x^2/4}}{1+\rme^{\delta x^2/4}}
 \right) \phi_\infty^\pm \nnl
&\bleq + \frac{\Gamma \rme^{-R^2}}{\sqrt{\pi} R^3}
\left[
\frac{\delta x}{1 + \rme^{-\delta x^2/4}} \pm \Delta x \left(\frac{1}{1 + \rme^{-\delta x^2/4}}
+ \sqrt{\pi} \, \delta x \right) \right] \,.
\end{align}
Since $\phi_\infty^\Sigma = 0$, we find
\begin{equation}\label{eqn:phase-large-r}
\phi_R^\Sigma \approx
\frac{\Gamma \rme^{-R^2} \delta x}{\sqrt{\pi} R^3 \left(
1 + \rme^{-\delta x^2/4}\right)} \,.
\end{equation}
Although the phase average vanishes exponentially for large $R$, it yields nonzero values for any finite value of $R$.

\section{Witnesses for spin entanglement}\label{sec:witness}

\begin{figure}
	\centering
	\includegraphics[scale=0.6]{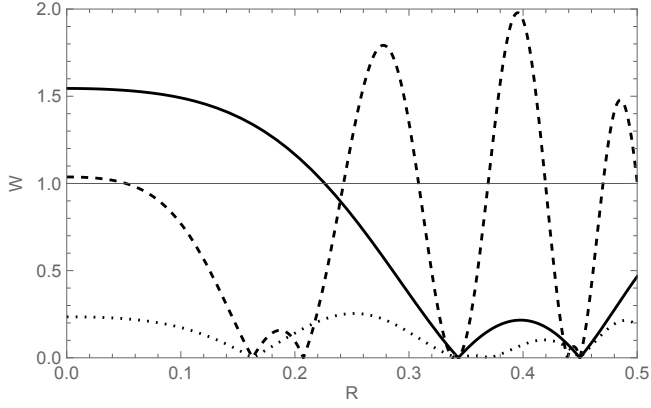}
	\caption{Entanglement witness $W$ defined in equation~\eqref{eqn:witness_fun_res} as a function of $R$ for the wide wave function case $\Delta x=0.25$, $\delta x=0.1$ and with values for $\Gamma$ of $0.5, 1, 2$.
	\label{fig:witness-plots}}
\end{figure}

With the phases derived in the previous section, we can now turn towards the task of witnessing the gravitationally induced entanglement experimentally. Considering the double Stern--Gerlach experiment described in section~\ref{sec:sg} and depicted in figure~\ref{fig:superpositions}, we first notice that the entanglement in the position degree of freedom gets transferred to the spin wave function of the whole system, which after passing through the interferometer and factoring out the global phase $\Phi_R$ reads
\begin{equation}\label{eqn:wf_spin}
	\ket{\psi} = \frac{\rme^{\rmi \Phi_R}}{2} \left(
	\ket{\uparrow\uparrow}
	+ \rme^{\rmi \phi_R^+} \ket{\uparrow\downarrow}
	+ \rme^{\rmi \phi_R^-} \ket{\downarrow\uparrow}
	+ \ket{\downarrow\downarrow}
	\right) \,,
\end{equation}
with the phases $\phi_R^\pm$ given by equation~\eqref{eqn:phi-plusminus}.
From the state~\eqref{eqn:wf_spin} we can calculate expectation values for chosen spin observables and their correlations. Bose et al.~\cite{boseSpinEntanglementWitness2017b} propose to witness spin entanglement via the function
\begin{equation}\label{eqn:witness_func}
	W = \abs{\langle \sigma_x^{(1)} \otimes \sigma_z^{(2)} \rangle + \langle \sigma_y^{(1)} \otimes \sigma_y^{(2)} \rangle} \,,
\end{equation}
where $\sigma_{x,y,z}^{(i)}$ denote the Pauli matrices acting on the spin of the $i$-th particle.
Writing explicitly
\begin{equation*}
	\sigma_x^{(1)} \otimes \sigma_z^{(2)} = 
	\begin{pmatrix}
		0 & 0 & 1 & 0 \\
		0 & 0 & 0 & -1 \\
		1 & 0 & 0 & 0 \\
		0 & -1 & 0 & 0 \\
	\end{pmatrix}
	,\quad
	\sigma_y^{(1)} \otimes \sigma_y^{(2)} = 
	\begin{pmatrix}
		0 & 0 & 0 & -1 \\
		0 & 0 & 1 & 0 \\
		0 & 1 & 0 & 0 \\
		-1 & 0 & 0 & 0 \\
	\end{pmatrix}
\end{equation*}
in the basis $\{\ket{\uparrow\uparrow}, \ket{\uparrow\downarrow}, \ket{\downarrow\uparrow}, \ket{\downarrow\downarrow}\}$ and with the state~\eqref{eqn:wf_spin}, this witness function evaluates to
\begin{align}\label{eqn:witness_fun_res}
	W &= \frac{1}{2} \abs{ \cos \left(\phi_R^-\right)+\cos \left(\phi_R^- -\phi_R^+ \right) - \cos \left(\phi_R^+ \right)-1} \nnl 
	&= \abs{\sin \phi_R^\Delta \left(\sin \phi_R^\Delta - \sin \phi_R^\Sigma\right)}\,.
\end{align}
This equation shows the dependence on the parameter $R$ via the phase $\phi_R^\pm$. Having learned, both from the plots in figure~\ref{fig:witness-plots} and the explicit form~\eqref{eqn:phase-large-r}, that in the limit $R \to \infty$ of mean-field semiclassical gravity one has $\phi_\infty^+ = -\phi_\infty^-$, with the symmetry of the cosine, one can see immediately that the witness function can only take values $0 \leq W \leq 1$, depending on the phase difference $\phi_\infty^\Delta$. It is also straightforward to show, that $W \leq 1$ for any separable spin wave function, implying that it witnesses entanglement for any value $W > 1$.

Since we can freely fix the phase difference $\phi_R^\Delta$ by adjusting the constant $\Gamma$, equation~\eqref{eqn:witness_fun_res} suggests that for any combination of phases with an average phase $\phi_R^\Sigma > 0$ entanglement witnessing values of $W > 1$ are possible, which is the case for any finite $R$. This is confirmed by the plot in figure~\ref{fig:witness-plots}, where the value of the entanglement witness is shown for different choices of the constant $\Gamma$ for small $R$. For large $R$, we can approximate and expand around $\phi_\infty^\Delta = -\pi/2$ in order to obtain
\begin{align}
W_\text{large $R$}
&\approx \abs{1 + \frac{\sqrt{\pi}}{2} R \rme^{-R^2} \delta x^2
  \frac{1+2 \rme^{\delta x^2/4}}{1+\rme^{\delta x^2/4}}
  \left(\phi_\infty^\Delta + \frac{\pi}{2}\right)}\,,
\end{align} 
which exceeds unity for any $\phi_\infty^\Delta > -\pi/2$. For large $\Delta x = 2 \delta x$, we find approximately $\phi_\infty^\Delta \approx -\Gamma/(3 \delta x)$, i.\,e. we can achieve entanglement for $2 \Gamma \approx 3 \pi \delta x$. However, in order to obtain an observable phase, the distances $\Delta x = 2 \delta x$, and thus also $\Gamma$, must grow exponentially with $R^2$.

Hence, we can claim that our class of deterministic models with a classical gravitational interaction predicts entanglement for an experimental setup such as the considered one, except in the strict limit $R \to \infty$, although the entanglement decreases rapidly with increasing $R$.

\begin{figure}
	\centering
	\includegraphics[scale=0.55]{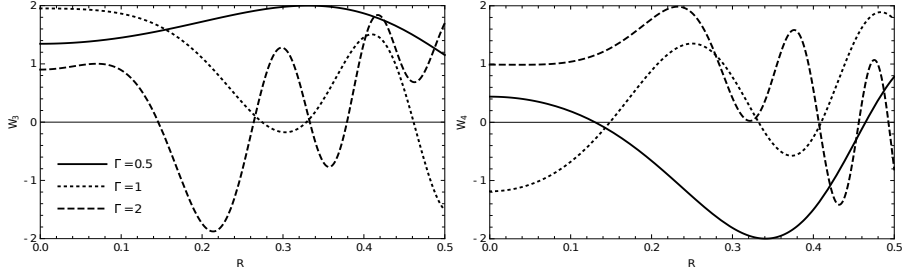}
	\caption{Change of special cases: $W_3 = W_G(\frac{3\pi}{2})$; small and $W_4 = W_G(\frac{\pi}{2})$; maximally entangled cases of the system, by the values of R with respect to the chosen numbers of $\Gamma$; $0.5, 1, 2$ and $R_{max}=0.5$ with the help of equation~\eqref{eqn:glo_witness_func}}
	\label{fig:gwitness3_4-plot}
\end{figure}

A more general treatment of spin entanglement witnesses has been presented by Guff et al.~\cite{guffOptimalFidelityWitnesses2021}. Generalizing the state~\eqref{eqn:wf_spin} to
\begin{equation}
	\ket{\psi(\alpha,\beta)} = \frac{1}{2} \left(
	\ket{\uparrow\uparrow}
	+ \rme^{\rmi \alpha} \ket{\uparrow\downarrow}
	+ \rme^{\rmi \beta} \ket{\downarrow\uparrow}
	+ \ket{\downarrow\downarrow}
	\right) \,,
\end{equation}
one can introduce a class of witness functions $W_G(\theta)$, parameterized by $\theta$ and defined via the projection on the state $\ket{\theta}=\frac{1}{\sqrt{2}}\left(\ket{\psi(0,0)}+\rme^{\rmi\theta}\ket{\psi(\pi,\pi)}\right)$:
\begin{align}\label{eqn:glo_witness_func}
	W_G(\theta) &= \ev{2 I- 4 \ket{\theta}\bra{\theta}}
	\nnl &= \langle I^{(1)} \otimes I^{(2)} \rangle - \langle \sigma_x^{(1)} \otimes \sigma_x^{(2)} \rangle + \cos(\theta)\left(\langle \sigma_y^{(1)} \otimes \sigma_y^{(2)} \rangle - \langle \sigma_z^{(1)} \otimes \sigma_z^{(2)} \rangle\right) 
	\nnl &\bleq + \sin(\theta)\left( \langle \sigma_y^{(1)} \otimes \sigma_z^{(2)} \rangle + \langle \sigma_z^{(1)} \otimes \sigma_y^{(2)} \rangle\right) \,,
\end{align}
with $I$ denoting the identity operator. This witness is scaled differently from $W$ above, with negative values indicating entanglement.
Evaluating the expectation values in the state~\eqref{eqn:wf_spin}, one finds
\begin{equation}
	W_G(\theta) = 2 \sin ^2\left(\frac{\theta}{2}\right) \sin ^2 \phi_R^\Delta + \sin \theta \left( \sin\phi _R^- + \sin\phi _R^+ \right) \,.
\end{equation}
A large area of the two-dimensional parameter space spanned by the phases $\phi_R^+$ and $\phi_R^-$ can be covered with only the witnesses $W_3 = W_G(\frac{3\pi}{2})$ and $W_4 = W_G(\frac{\pi}{2})$~\cite{guffOptimalFidelityWitnesses2021}, which can be seen as witness functions optimized for the detection of small entanglement and maximally entangled states of the system, respectively. Plots of these witnesses for different values of the parameter $R$ are shown in figure~\ref{fig:gwitness3_4-plot}, again confirming that entanglement can be observed for finite values of $R$.

Because we can always choose parameters $m$ and $\tau$ such that $\phi_R^\Delta$ is a multiple of $\pi$, $W_G(\theta)$ can then always take negative values \emph{unless} $\phi_R^\Sigma$ is also a multiple of $\pi$. Specifically, no entanglement can be observed in the limit $R \to \infty$ where $\phi_R^\Sigma \to 0$, whereas for large but finite $R$ one finds $0 < \abs{\phi_R^\Sigma} \ll \pi$ and negative values of $W_G(\theta)$ are possible at least in principle, if decoherence effects can be kept small.

\section{Discussion}\label{sec:discussion}

We presented a class of models for nonrelativistic quantum systems on a classical spacetime, where the curvature of spacetime is sourced in a semiclassical fashion, depending on both the wave function and the particle trajectory in the sense of de Broglie--Bohm theory. Except for the limiting case $R \to \infty$, where our models yield the pure mean-field semiclassical gravity model based on the semiclassical Einstein equations~\eqref{eqn:sce}, all models in this class have the capability to generate entanglement between two particles.

We are not making any claim for these models to be a realistic representation of how gravity works in the regime of nonrelativistic quantum systems. However, they show as a proof of principle that there \emph{can} be models that i.) result in entanglement between two particles, ii.) allow for an interpretation as the nonrelativistic limit of a theory for quantum matter on a classical spacetime, and iii.) are physically inequivalent to standard quantum mechanics, i.\,e. the nonrelativistic limit of perturbative quantum gravity in analogy to the limit of quantum electrodynamics to the Coulomb potential. In this regard, our models provide explicit counterexamples to the arguments that experimental evidence of entanglement would prove the necessity to quantize the gravitational field. Note that by ``quantized'' we mean the impossibility to describe spacetime as a classical Lorentzian 4-manifold. One could, of course, adopt a different notion of quantumness in which entanglement is a defining feature; this, however, would render the argument that entanglement provides evidence for quantization tautological.

An important caveat concerns the function $\gamma_R$ we added to the potentials~\eqref{eqn:vbb} and~\eqref{eqn:vr}. Without this function, the semiclassical interpretation of the gravitational field as being sourced by the mass $m$ distributed with the modulus squared of the wave function over a radius $R$ around the Bohmian positions is evident, and it provides the desired counterexample. If, however, $R$ is below the size of superpositions for which the gravitational phase shift~\cite{colellaObservationGravitationallyInduced1975,nesvizhevskyMeasurementQuantumStates2003,fixlerAtomInterferometerMeasurement2007} has been observed, a nonzero $\gamma_R$ is required for consistency with these observations. The interpretation of the gravitational force as a consequence of a single, classical spacetime then becomes less convincing.

Note also that the conclusion from our discussion here is \emph{not} that any of the theorems regarding entanglement via classical and nonclassical channels, e.\,g. in reference~\cite{marlettoWitnessingNonclassicalityQuantum2020}, are incorrect. Rather our models show that for the purpose of exploring semiclassical alternatives to quantum gravity, the assumptions underlying these theorems could be too constraining.
In this context, it is interesting to have a closer look at the one assumption stated explicitly by Marletto and Vedral~\cite{marlettoAnswersFewQuestions2019}, namely the locality assumption ``that the two objects to be entangled should not be interacting directly, but only locally, at their respective locations, with the mediator.''~\cite{marlettoAnswersFewQuestions2019}
The mediator of the gravitational interaction is spacetime curvature, with which the wave function interacts entirely locally via the Newtonian potential, exactly like in standard quantum mechanics.
There is, however, a different nonlocal element in our models---as it must be in order to account for the violation of Bell's inequalities---which is the dependence of the \emph{source} of spacetime curvature not only on the local value of the wave function (as in the mean-field approach) but also on the Bohmian trajectories.

In this context, it is important that the models defined in sections~\ref{sec:bohmian} and~\ref{sec:interpol} are nonrelativistic, and perfectly consistent as such. The attempt to find a relativistic version is met with difficulties, as the particle coordinates---or their field theoretic complements---in the de Broglie--Bohm theory do not conserve energy~\cite{struyveSemiclassicalApproximationsBased2020}. Nonetheless, the proposed experimental tests for entanglement are only formulated nonrelativistically themselves. To this effect, one must keep in mind that not only are the semiclassical Einstein equations equally inconsistent if not endowed with some objective collapse mechanism~\cite{pageIndirectEvidenceQuantum1981}; even the quantum potential~\eqref{eqn:vqg} is the nonrelativistic limit of a theory (perturbative quantum gravity) known to be non-renormalizable at high energies. The question whether for a given nonrelativistic potential there is \emph{any} complete and consistent relativistic theory that limits to said potential should, therefore, be considered an open one for both semiclassical \emph{and} quantized gravity.

Experiments in physics ultimately serve two purposes. On one hand, they can increase our trust in the established theoretical frameworks by confirming their predictions. An experimental confirmation of gravitational entanglement could considerably increase our confidence in perturbative quantum gravity as the low-energy limit of whatever the correct quantum theory of gravity may be. A failure to demonstrate entanglement, by contrast, would create serious doubt about traditional approaches. In this sense, tests for gravitationally induced entanglement are an invaluable tool.
On the other hand, experiments can never give \emph{proof} of any particular model. What they can do instead, is rule out certain elements from a set of plausible alternative models. If one puts the bar for theories describing the gravitational interaction of quantum systems as high as only allowing fully consistent relativistic models into this set of possibilities, one is effectively ruling out elements from an empty set---and all experiments seem equally useless. In order to arrive at meaningful statements about what is truly known \emph{empirically} about quantum gravity, one should allow for candidate models to possess limitations, and carefully distinguish \emph{inconsistency} in a strict mathematical sense from mere \emph{incompleteness} for which it cannot be conclusively ruled out that there might be a mathematically consistent way---as implausible as it may appear---towards a full theory.
Our point of view is that there are models that fall into the latter category and predict entanglement through interaction with a classical spacetime.

%
%

\begin{acknowledgements}
The authors gratefully acknowledge funding by the Volkswagen Foundation.
\end{acknowledgements}

\section*{Appendix}
\begin{appendix}
Let $\Psi : \reals^{1+N} \to \complex$, $\vec q : \reals \to \reals^N$ be a solution of the coupled system
\begin{align}
\rmi \hbar \frac{\partial}{\partial t} \Psi(t,\vec x)
&= - \sum_{i=1}^N \frac{\hbar^2}{2 m_i} \frac{\partial^2}{\partial x_i^2} \Psi(t,\vec x)
+ V[\Psi](t,\vec x,\vec q) \Psi(t,\vec x) \label{eqn:app-psi}\\
\frac{\rmd}{\rmd t} q_i(t) &= \frac{\hbar}{m_i} \left.\Im\left(\frac{\partial \Psi(t,\vec x)/\partial x_i}{\Psi(t,\vec x)}\right)\right\vert_{\vec x = \vec q(t)} \,, \label{eqn:app-q}
\end{align}
and let $\Psi_f : \reals^{1+N} \to \complex$ be the solution of the \emph{free} \schr\ equation
\begin{equation}
\rmi \hbar \frac{\partial}{\partial t} \Psi_f(t,\vec x)
= - \sum_{i=1}^N \frac{\hbar^2}{2 m_i} \frac{\partial^2}{\partial x_i^2} \Psi_f(t,\vec x) \label{eqn:app-psifree}
\end{equation}
with the same initial conditions, $\Psi_f(0,\vec x) = \Psi(0,\vec x)$.
Define the free classical trajectories
\begin{equation}
u_i(t) = \int \rmd^N x \, x_i \abs{\Psi_f(t,\vec x)}^2
\end{equation}
and the phase function $S : \reals \to \reals$ by
\begin{equation}
S(t) = -\frac{1}{\hbar} \int_0^t \rmd t' \, V[\Psi_f](t',\vec u,\vec u) \,.
\end{equation}
Furthermore, we define
\begin{align}
 \delta \Psi(t,\vec x) &= \Psi(t,\vec x) - \rme^{\rmi S(t)} \Psi_f(t,\vec x) \\
 \delta V(t,\vec x) &= V[\Psi](t,\vec x,\vec q) - V[\Psi_f](t,\vec u,\vec u)
\end{align}
which satisfy the \schr\ equation
\begin{equation}
 \rmi \hbar \frac{\partial}{\partial t} \delta\Psi(t,\vec x) =
 - \sum_{i=1}^N \frac{\hbar^2}{2 m_i} \frac{\partial^2}{\partial x_i^2} \delta\Psi(t,\vec x)
+ V[\Psi](t,\vec x,\vec q) \delta\Psi(t,\vec x) 
+ \delta V(t,\vec x)
\rme^{\rmi S(t)}\Psi_f(t,\vec x) \,.
\end{equation}
Assume that $\Psi_f(t,\vec x)$ remains confined to an open neighborhood $\Sigma_t \subset \reals^N$ of $\vec u(t)$ for any time $t \in [0,T]$, i.\,e. there is an $\epsilon_1 > 0$ such that
\begin{equation}
\forall t \in [0,T],\;\forall \vec x \in \reals^N \backslash \Sigma_t : \quad
 \abs{\Psi_f(t,\vec x)} < \epsilon_1 \,.
\end{equation}
As a consequence of the quantum equilibrium hypothesis~\cite{durrBohmianMechanicsPhysics2009}, which states that the Bohmian trajectories $\vec q$ are distributed with $\abs{\Psi(0,\vec x)}^2$, one expects them to remain close to the classical ones and we can approximate $\vec u \approx \vec q$.
Further assume that within each $\Sigma_t$ the potential changes sufficiently slowly with position and wave function, i.\,e. there is an $\epsilon_2 > 0$ such that
\begin{equation}
\forall t \in [0,T],\;\forall \vec x \in \Sigma_t : \quad
 \abs{\delta V(t,\vec x)} < \epsilon_2 \,.
\end{equation}
The $L_2$ norm of $\delta \Psi$ then satisfies, using the Cauchy-Schwarz inequality,
\begin{align}
 \abs{\partial_t \norm{\delta \Psi}_2^2} &= 2 \norm{\delta \Psi}_2 \abs{\partial_t \norm{\delta \Psi}_2} \nnl &=
 \abs{\int \rmd^N x \, \left( \delta\Psi^*(t,\vec x) \frac{\partial}{\partial t} \delta\Psi(t,\vec x)
 + \delta\Psi(t,\vec x) \frac{\partial}{\partial t} \delta\Psi^*(t,\vec x) \right)} \nnl
 &= \frac{2}{\hbar} \abs{\int \rmd^N x\, \delta V(t,\vec x) \Im\left(
 \rme^{\rmi S(t)} \delta \Psi^*(t,\vec x) \Psi_f(t,\vec x) \right)}
 \nnl &\leq \frac{2}{\hbar} \left( \int_{\Sigma_t} \rmd^N x\, \abs{\delta V(t,\vec x)} \abs{\delta \Psi(t,\vec x)} \abs{\Psi_f(t,\vec x)}
 + \int_{\reals^N \backslash \Sigma_t} \rmd^N x\, \abs{\delta V(t,\vec x)} \abs{\delta \Psi(t,\vec x)} \abs{\Psi_f(t,\vec x)} \right)
 \nnl &\leq \frac{2}{\hbar} \left( \epsilon_2 \int \rmd^N x\, \abs{\delta \Psi(t,\vec x)} \abs{\Psi_f(t,\vec x)}
 + \epsilon_1 \int \rmd^N x\, \abs{\delta V(t,\vec x)} \abs{\delta \Psi(t,\vec x)} \right)
 \nnl &\leq \frac{2}{\hbar} \left( \epsilon_2 \norm{\delta \Psi}_2 \norm{\Psi_f}_2
 + \epsilon_1 \norm{\delta V}_2 \norm{\delta \Psi}_2 \right)
 \nnl &\leq \frac{2}{\hbar} \left( \epsilon_2  
 + \epsilon_1 \norm{\delta V}_2 \right)
 \norm{\delta \Psi}_2 \,,
\end{align}
and we find
\begin{equation}
\forall t \in [0,T] : \quad
\norm{\delta \Psi}_2 \leq T \, \max_{t \in [0,T]}
 \abs{\partial_t \norm{\delta \Psi}_2}
 \leq \frac{T}{\hbar} \left(\epsilon_2  
 + \epsilon_1 \norm{\delta V}_2\right) \,.
\end{equation}
Hence, $\Psi$ is well approximated by $\rme^{\rmi S(t)} \Psi_f$ for at least some time $T$, as long as $\Psi_f$ is sufficiently constrained around the classical trajectory $\vec u(t)$ and the potential depends slowly on position and reacts sufficiently slowly to changes in the wave function.

If the system under consideration is initially in a superposition state $\Psi(0,\vec x) = \Psi_f(0,\vec x) = \sum \alpha_j \Psi_j(0,\vec x)$, and the \schr\ equation~\eqref{eqn:app-psi} is linear, then the previous considerations hold for every branch $\Psi_j$ independently, which acquire independent phases $S_j(t)$. This is no longer true for a nonlinear potential $V[\Psi]$, where a mixing of branches can occur. However, due to the linearity of the free \schr\ equation~\eqref{eqn:app-psifree} no mixing occurs for the branches of $\Psi_f$. If, as for the gravitational potential~\eqref{eqn:vr}, $V[\Psi]$ only depends on the modulus and not the acquired phases $S_j(t)$ of the wave function, no mixing occurs in the approximation where $V[\Psi] \approx V[\Psi_f]$, which renders the \schr\ equation for $\Psi$ effectively linear. In this approximate sense, one can justify the description of the gravitational effect as introducing phases to each possible combination of classical trajectories.
\end{appendix}

%
\section*{Conflict of interest}
The authors declare that they have no conflict of interest.


\end{document}